\documentclass[useAMS,usenatbib]{mn2e}
\usepackage{graphicx}
\usepackage{rotating}
\usepackage{longtable}
\usepackage{lscape}
\usepackage{amssymb,amsmath}
\def\lsim{\mathrel{\rlap{\lower 3pt \hbox{$\sim$}} \raise 2.0pt \hbox{$<$}}}
\def\gsim{\mathrel{\rlap{\lower 3pt \hbox{$\sim$}} \raise 2.0pt \hbox{$>$}}}
\def\degree{\ensuremath{^\circ}}
\def\msun{{\rm M}_\odot}

\def\mgii{{\rm Mg\,{\sc II}}}

\def\feii{{\rm Fe\,{\sc II}}}

\def\dv{\Delta{\rm V}}
\def\ad{\Delta\theta}
\def\pd{{\rm pd}}
\def\z{{\rm z}}
\def\kms{km\,s$^{-1}$}
\def\qqq{QQQ~J1519+0627}
\def\qqqa{QQQ1519A}
\def\qqqb{QQQ1519B}
\def\qqqc{QQQ1519C}

\title[Discovery of a Physical Quasar Triplet]
	{Caught in the Act: \\ Discovery of a Physical Quasar Triplet}
\author[Farina et al.]{
	E.~P.~Farina$^{1,2}$\thanks{E--mail: {\tt emanuele.paolo.farina@gmail.com}},
	C.~Montuori$^{1,3}$, 
	R.~Decarli$^{4}$, and
	M.~Fumagalli$^{5,6}$\thanks{Hubble Fellow}\\
       	$^{1}$ Universit\`{a} degli Studi dell'Insubria, via Valleggio 11, I-22100 Como, Italy\\
	$^{2}$ INFN Milano--Bicocca --- Universit\`{a} degli Studi di Milano--Bicocca, Piazza della Scienza 3, I-20126 Milano, Italy\\
	$^{3}$ Technion, Department of Physics, IL-32000 Haifa, Israel\\
	$^{4}$ Max-Planck-Institut f\"ur Astronomie, K\"onigstuhl 17, D-69117 Heidelberg, Germany\\
       	$^{5}$ Carnegie Observatories, 813 Santa Barbara Street, CA-91101 Pasadena, USA\\
	$^{6}$ Department of Astrophysics, Princeton University, NJ-08544-1001 Princeton, USA
	}
\begin{document}

\date{}

\pagerange{\pageref{firstpage}--\pageref{lastpage}} \pubyear{2013}

\maketitle

\label{firstpage}

\begin{abstract}
We present the discovery of a triplet of quasars at $\z\approx1.51$. The 
whole system is well accommodated within $25\arcsec$ (i.e., 200\,kpc in
projected distance). The velocity differences among the three objects
(as measured through the broad \mgii\ emission line) are~$<1000$\,\kms,
suggesting that the quasars belong to the same physical structure. 
Broad band NIR images of the field do not reveal evidence of galaxies or 
galaxy clusters that could act as a gravitational lens, ruling out the 
possibility that two or all the three quasars are multiple images 
of a single, strongly lensed source. \qqq\ is the second triplet of 
quasars known up to date. We estimate that these systems are extremely 
rare in terms of simple accidental superposition. The lack of
strong galaxy overdensity suggests that this peculiar system is
harboured in the seeds of a yet--to--be--formed massive structure.\\

Based on observations collected at the La Silla Observatory with the New 
Technology Telescope (NTT) of  the European Southern Observatory (ESO)
and at the Calar Alto Observatory with the 3.5m telescope of the Centro 
Astr{\'o}nmico Hispano Alem{\'a}n (CAHA). 
\end{abstract}


\section{Introduction}\label{sec:1}

In the current hierarchical cosmological paradigm, galaxy mergers and 
interactions are a major route to galaxy formation. The strong tidal 
fields experienced during close encounters between gas rich systems 
trigger large--scale nuclear inflows that can ultimately start the 
activity of the supermassive black holes (SMBHs) residing in the central 
regions of the interacting systems \citep[e.g.,][]{DiMatteo2005}. 
The existence of multiple, simultaneously active, SMBHs represents a 
key observational test for this evolutionary scenario and for our 
understanding of the processes regulating the quasar (QSO) activity 
and the co--evolution of SMBHs with their host galaxies.

A number of studies demonstrated the connection between QSOs and 
ultra--luminous infrared galaxies that are mostly found in merging systems
\citep[e.g.,][]{Canalizo2001} and confirmed the signature of recent
merger events in QSO hosts \citep[e.g.,][]{Bennert2008}. A noticeable
example is the discovery of a spatially resolved binary QSO 
(projected separation $\pd=21$\,kpc) clearly hosted by a galaxy merger 
\citep{Green2010}. 
The search for candidate binary systems of QSOs up to redshift $\sim4$ 
unveiled an enhanced QSO clustering signature at small scale 
($\lesssim50$\,kpc) relative to the simple extrapolation of the 
larger scale two--point correlation function (e.g., 
\citealp{Hennawi2006, Hennawi2010, Myers2007, Myers2008, Kayo2012}; 
but see \citealp{Padmanabhan2009, Shen2010}). 
This can be interpreted in terms of interactions that
trigger the QSO activity \citep[e.g.,][]{Hennawi2006, Hopkins2008}. 
Alternatively, the small scale excess could be a simple manifestation 
of the clustering properties of the halos that host QSOs 
\citep[e.g.,][]{Richardson2012}.

Little is known about systems with more than two physically associated
QSOs which are expected to be even more elusive objects. The only
physical QSO triplet reported so far is QQQ~J1432-0106 at $\z=2.076$ 
observed by \citet{Djorgovski2007}.  
At smaller projected separations and lower luminosities, \citet{Liu2011}
recently discovered a triple AGN in the galaxy SDSS~J1027+1749
at $\z=0.066$, and \citet{Schawinski2011} serendipitously observed three 
low mass ($<10^7\,\msun$) accreting black holes in a galaxy at $\z=1.35$.

In this Paper we present the discovery of \qqq, the second triplet of 
QSOs that is known up to date. This is the first result of our ongoing 
systematic search for QSO triplets among the photometric and 
spectroscopic database of the Sloan Digital Sky Survey 
\citep[SDSS;][]{Aihara2011}. The manuscript is organised as 
follows. In~Section~\ref{sec:2} we describe the spectroscopic and photometric 
data collected in our study. The possible interpretations are discussed
in~Section~\ref{sec:3}, and we draw our conclusions 
in~Section~\ref{sec:4}.

Throughout this paper we consider a concordance cosmology with 
H$_0=70$\,\kms\,Mpc$^{-1}$, $\Omega_{\rm m} = 0.3$, and $\Omega_\Lambda=0.7$.
All the quoted magnitude are expressed in the AB standard photometric 
system \citep{Oke1974}.

\section[]{Selection and Follow--up Observations}\label{sec:2}

QSO multiplets observed at close projected separations are rare 
\citep[e.g.,][]{Hennawi2006}. Large spectroscopic surveys often 
fail to detect close QSO systems due to the fiber collision limits. 
For example in SDSS it is not possible to obtain the spectrum for 
both sources in a pair with separation $<55\arcsec$ within a single 
plate \citep{Blanton2003}. To overcome this limitation, we started a 
programme to search for close QSO triplets taking advantage of the 
large photometric sample of \citet{Richards2009}. 
Three QSOs are considered a candidate triplet if (i) at least one 
of them has spectroscopic redshift, (ii) the other two reside within 
500\,kpc from it, and (iii) they have coincident (within the 
uncertainties) photometric redshift \citep[estimated with the procedure 
suggested by][]{Weinstein2004}. 
We selected 13 triple QSO {\it candidates} for spectroscopic 
follow up, in the redshift range $0.3\lesssim \z \lesssim 2.2$, 
and with an average projected separation of $\sim300$\,kpc.
Further details on the selection procedure will be provided in a 
forthcoming paper (Farina~et~al., in preparation).
Here we present the spectroscopic observations obtained for the first
followed--up target among our candidates. In particular, we report 
our discovery of two QSOs within~$\sim25\arcsec$ from the 
spectroscopic QSO SDSS~J151947.3+062753 
\citep[hereafter \qqqa,][]{Schneider2010} and located at a similar
redshift (radial velocity difference \mbox{$\dv\lsim1000$\,\kms}). 

To provide a rough estimate of the expected number of such
systems, we derived the QSO three--point correlation 
function ($\zeta$) from the amplitude of the two--point correlation 
function ($\xi$): 
$\zeta_{123}(r)=\mathcal{Q}\,\left[\xi_{12}(r)+\xi_{23}(r)+\xi_{31}(r)\right]$,
where $\mathcal{Q}\approx1$ \citep[e.g.,][]{Peebles1993, Djorgovski2007}.
Assuming the values of the projected two--point correlation function
from \citet{Hennawi2006} and integrating the QSO luminosity function
presented by \citet{Croom2009} above the magnitude range spanned by 
the photometric selected QSOs (i.e., from ${\rm M}_{\rm g}\approx-28.0$
to $\approx-25.5$) we calculated that in a perfect survey, with no 
source of incompleteness, given a QSO, the probability of finding two 
companions within $500$\,kpc is $p_{{\rm ABC}}\approx10^{-8}$. 
Out of the \citet{Schneider2010} catalogue, which consists in
$\sim106000$ spectroscopically confirmed QSOs in the SDSS
Data Release 7 footprint, we thus expect that $\sim0.002$ objects
have two companions. The mere observation of one (or more) systems
of this kind would substantially strengthen the argument of 
small--scale enhancement of QSO clustering \citep[e.g.,][]{Myers2007}.

In Table~\ref{tab:triple} we list the properties of the QSOs belonging to 
our newly discovered triplet, labelled \qqq, and, for comparison,
of the components of the only other triple QSO system known so far: 
QQQ~J1432-0106. 
In the following paragraphs we describe the procedures and the results 
of the analysis of the spectroscopic and photometric data collected 
on \qqq. 

\begin{table*}
\centering
\caption{
Properties of the triple QSO systems known to date.
For each QSO, we list:
identification label (id),
position (RA, DEC),
redshift (redshift),
magnitude of the components (z, J, and H),
angular ($\ad$) and
projected ($\pd$) separations at the redshift of the system,
and radial velocity difference ($\dv$).
Data for QQQ~J1432-0106 are from \citet{Djorgovski2007}
and the SDSS database.
}\label{tab:triple}
\begin{tabular}{lccccccccc}
\hline
id	        & RA  	     & DEC	  & redshift        & z     & J     & H     & $\ad$  	                & \pd			 & $\dv$ 		     \\
                & [J2000]    & [J2000]    &		    & [mag] & [mag] & [mag] & [arcsec]		        & [kpc] 		 & [\kms]		     \\
\hline  		 
\hline  		 
\qqq:           &            &            &                 &       &       &       &                           &			 &			     \\
 -- \qqqa       & 15:19:47.3 &  +06:27:53 & 1.504$\pm$0.001 & 18.86 & 18.81 & 18.55 & $\ad$(A-B)=23.5	        & \pd(A-B)=198		 & $\dv$(A-B)=1100  	     \\
 -- \qqqb       & 15:19:45.7 &  +06:27:52 & 1.513$\pm$0.003 & 21.23 & 20.97 & 20.25 & $\ad$(B-C)=\phantom{2}3.7 & \pd(B-C)=\phantom{1}31 & $\dv$(B-C)=\phantom{1}850 \\
 -- \qqqc       & 15:19:45.9 &  +06:27:49 & 1.506$\pm$0.003 & 21.20 & 20.69 & 20.13 & $\ad$(C-A)=21.1	        & \pd(C-A)=178		 & $\dv$(C-A)=\phantom{1}250 \\
\hline
QQQ~J1432-0106: &            &            &                 &       &       &       &                           &		         &			     \\
 -- QQQ1432A    & 14:32:29.2 & --01:06:16 & 2.076	    & 17.15 & 16.32 & 15.88 & $\ad$(A-B)=\phantom{2}5.1 & \pd(A-B)=\phantom{1}42 & $\dv$(A-B)=\phantom{1}280 \\
 -- QQQ1432B    & 14:32:28.9 & --01:06:13 & 2.076	    & 20.42 & 20.00 & 19.27 & $\ad$(B-C)=\phantom{2}3.6 & \pd(B-C)=\phantom{1}30 & $\dv$(B-C)=\phantom{1}100 \\
 -- QQQ1432C    & 14:32:29.2 & --01:06:12 & $\sim$2.08      & \dots & 21.66 & 21.17 & $\ad$(C-A)=\phantom{2}4.3 & \pd(C-A)=\phantom{1}36 & \dots  	             \\
\hline
\end{tabular}
\end{table*}

\subsection[]{Long--slit spectroscopy}

Spectra of \qqqb\ and \qqqc\ were gathered with the ESO Faint Object 
Spectrograph and Camera~2 \citep[EFOSC2;][]{Buzzoni1984} mounted on the
New Technology Telescope (NTT) in La Silla (Chile). We performed long--slit
spectroscopy on February 17, 2012, using a slit width of $1\arcsec$ and
setting the Position Angle to $-49.7\degree$, so that both the QSOs
were observed simultaneously. Grism \#16 was used in order to continuously
cover the wavelength range 6015 - 10320\,\AA\ with a spectral resolution
$\lambda/\Delta \lambda \approx 550$ as measured on the night sky emission
lines. Five frames of 900\,s were acquired, for a total exposure time
of 75\,min on source.
Standard \texttt{IRAF}\footnote{\texttt{IRAF} is distributed by the National 
Optical Astronomy Observatories, which are operated by the Association of 
Universities for Research in Astronomy, Inc., under cooperative agreement with 
the National Science Foundation.} tools were used in the data reduction 
(bias subtraction, flat--fielding, wavelength and flux calibration, spectra 
extraction). As flux calibrator we observed the spectrophotometric standard 
star Hiltner~600. Typical residuals in the wavelength calibration are 
$\sim1$\,\AA.

\begin{figure*}
\centering
\includegraphics[width=2.00\columnwidth]{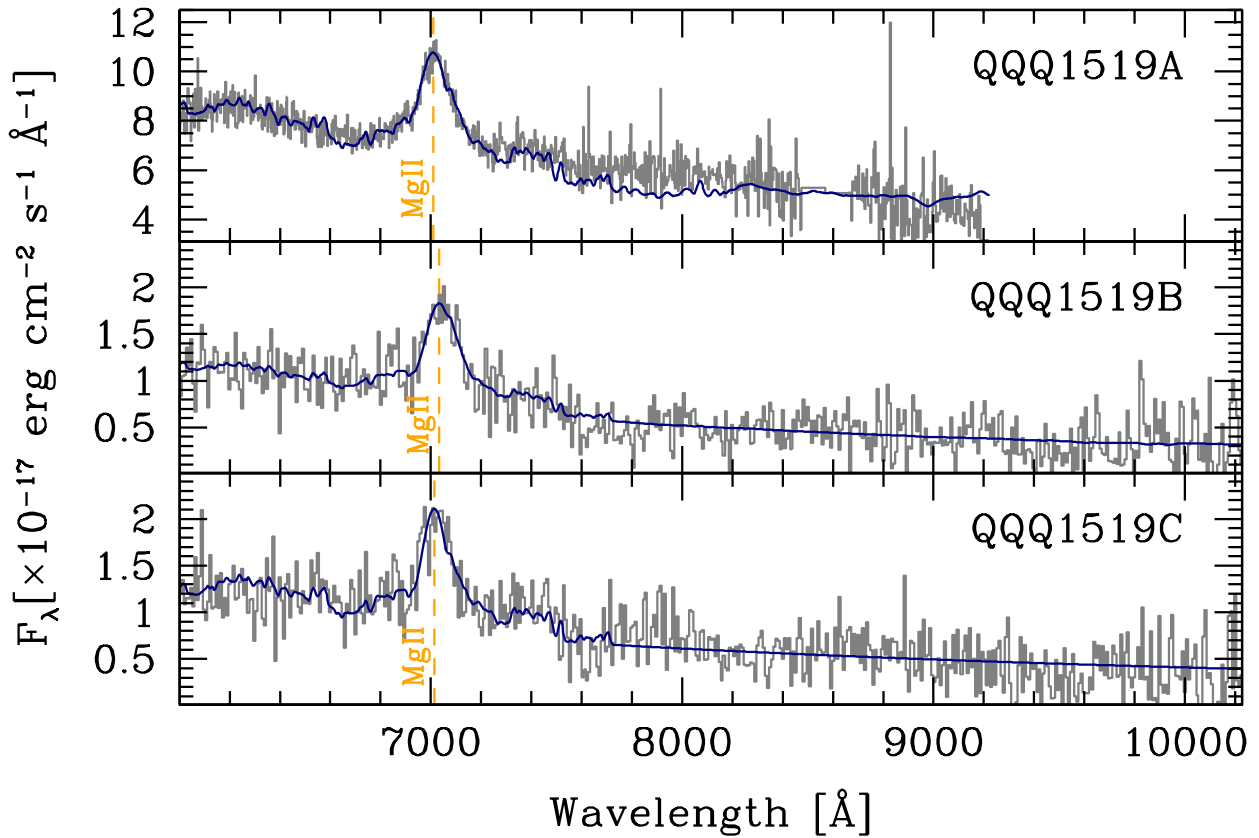}
\caption{Spectra of the three QSOs of \qqq\ (grey), binned by 2
pixels. The results of the fitting procedure on \mgii\ and \feii\ lines are 
plotted in blue (see text for details). The positions of the \mgii\ 
peaks are marked with orange dashed lines.
}\label{fig:1}
\end{figure*}

For the analysis of the QSO spectra we followed the procedure presented in 
\citet{Decarli2010b} and \citet{Derosa2011}. Namely, we model the QSO
spectra with a superposition of:
(i) a power--law non--thermal component;  
(ii) the emission of the host galaxy \citep[adopting the elliptical galaxy 
model by][]{Mannucci2001}; 
(iii) the \feii\ multiplets (from the template of \citealt{Vestergaard2001} 
and from our own original spectrum of IZw001), and
(iv) the broad emission lines fitted with two Gaussian curves with the same 
peak wavelength \citep[see][]{Decarli2008}. The two NTT spectra and the one
from SDSS are shown in Figure~\ref{fig:1}. A bright emission line,
identified as \mgii, is observed in all the spectra at $\sim7000$\,\AA. 
The line identification is supported by the lack of other bright
emission lines in the observed spectra, and by a tentative detection of
the iron multiplets around the line (see our fit results in 
Figure~\ref{fig:1}). These pieces of evidence firmly pin down the
redshifts of the QSOs to $1.504\pm0.001$, $1.513\pm0.003$ and 
$1.506\pm0.003$ for A,~B and~C respectively, where uncertainties are 
given by the positional accuracy of the  \mgii\ line centroids.

\subsection[]{Broad band photometry}\label{sec:2.2}

We gathered NIR broad band images of the \qqq\ field using Omega2000
\citep{Kovacs2004} at the 3.5\,m telescope in Calar Alto (Spain). 
Observations were performed on March 8, 2012 as Director Discretional 
Time. We observed a region of $15\arcmin\times15\arcmin$ around the 
system in z, J, and H bands, i.e., sampling the rest--frame
u, g, and r at $\z=1.51$. We adapted usual jittering observing strategies in order
to effectively subtract night sky emission. We collected $180 \times 15$\,s
frames in z (total integration time: 45\,min), $18 \times 100$\,s frames
in J (total integration time: 30\,min), and $36 \times 50$\,s frames in
H (total integration time: 30\,min). Images were processed with our 
package of NIR image processing based on standard \texttt{IRAF} tasks.
We compute the astrometric solution using the \texttt{astrometry.net}
software \citep{Lang2010}. Photometric calibration is achieved by 
comparing the photometry of field stars observed in our images with
the fluxes reported in the SDSS (z--band) and 2MASS (J-- and H--band) 
catalogues, using the following filter transformations:
\begin{eqnarray}
{\rm z}_{\Omega2k} &=& {\rm z}_{\rm SDSS}     - 0.05 \, {\rm (i-z)}_{\rm SDSS}      \label{eq_z} \\ 
{\rm J}_{\Omega2k} &=& {\rm J}_{\rm 2MASS,AB} + 0.11 \, {\rm (J-H)}_{\rm 2MASS,AB}  \label{eq_J} \\
{\rm H}_{\Omega2k} &=& {\rm H}_{\rm 2MASS,AB} - 0.02 \, {\rm (J-H)}_{\rm 2MASS,AB}  \label{eq_H}
\end{eqnarray}
Equations \ref{eq_z}--\ref{eq_H} are derived by computing the spectral 
magnitudes of template O-- to M--type stars.

The seeing during the observations was $2\arcsec$ (as measured on the final 
images), and the $5\,\sigma$ limit magnitudes (computed from the rms of sky 
counts over a seeing area) are 23.90, 23.38 and 22.25 in z, J and H 
respectively.

\begin{figure*}
\centering
\includegraphics[width=2.00\columnwidth]{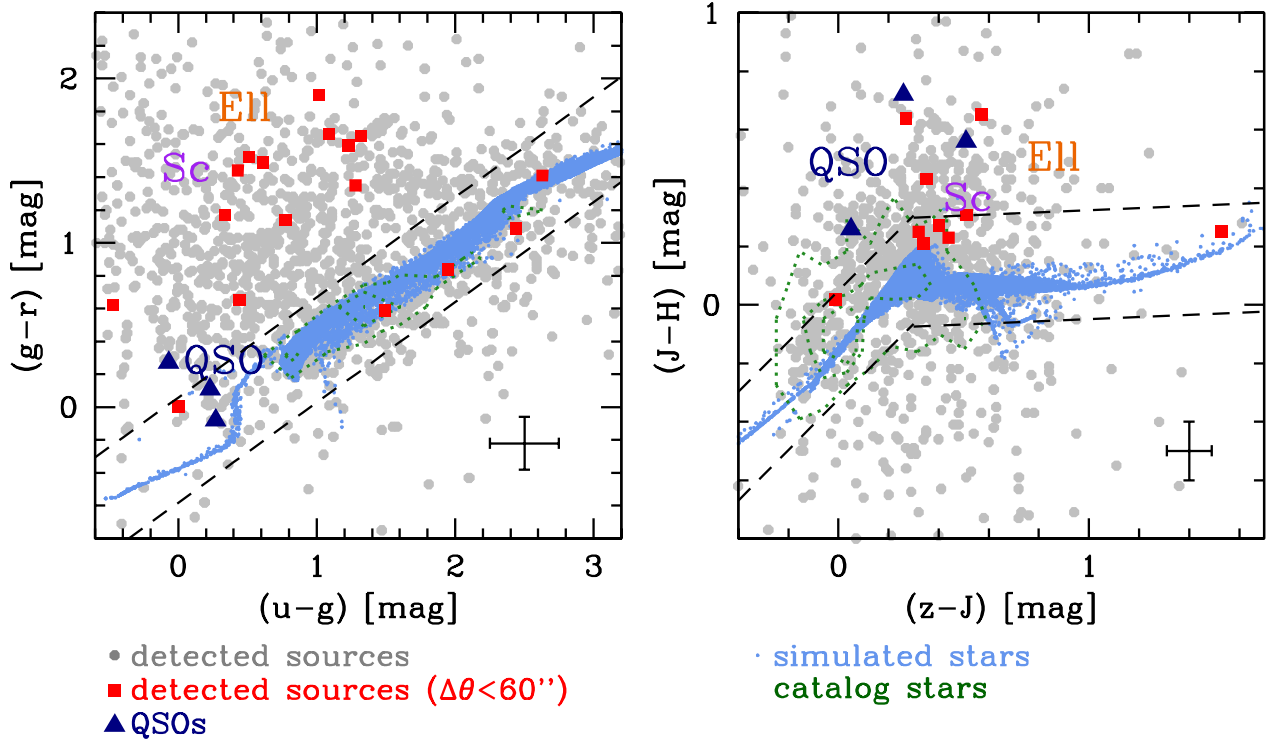}
\caption{Colour--colour diagrams of the sources detected in the explored
images (grey points). Objects within $60\arcsec$ from the \qqq\ baricenter 
are highlighted in red. The blue triangles indicate the colours of the 
three QSOs of the system. Note that the number of detected sources changes, 
according to the different depth of each band.
For comparison we show the expected loci of 
Ellipticals and Sc galaxies at $\z=1$ and of QSOs at $\z=1.5$ 
from the template of \citet{Mannucci2001} and \citet{Francis1991}.
The light blue points are the star colours 
simulated with the \texttt{Trilegal} software \citep{Girardi2005}, 
and dark green contours indicate the distribution of the 
spectroscopically confirmed stars detected in the SDSS and 2MASS  
in a region of $5\degr\times5\degr$ around the triplet, and in 
the GOODS survey.
Black dashed lines show the colour cuts we imposed to separate
stars and extragalactic objects. The crosses in the 
bottom--right of the figures represent typical error bars on 
the colours.}\label{fig:2}
\end{figure*}

We use \texttt{SExtractor} \citep{Bertin1996} in order to identify
and catalogue the sources in our images. Within the limit fluxes of
our observations, we are able to study the environment of the 
QSO system down to ${\rm M}^\star(\z)+1$, where ${\rm M}^\star(\z)$ is 
the rest--frame B--band characteristic luminosity of galaxies at redshift 
$\z$ as derived from the fit of the galaxy luminosity function \citep{Ilbert2005}.
In order to distangle between Galactic and extragalactic objects we rely upon 
different colour--colour diagrams, also including sources from the SDSS
photometric database (see Figure~\ref{fig:2}).

We classified a source as \textit{star} if it is not resolved (i.e., FWHM $\lsim$ seeing 
in all the considered images), and it lies within the locus of main sequence stars
estimated through the prediction of the \texttt{Trilegal} 
software\footnote{\texttt{http://stev.oapd.inaf.it/cgi-bin/trilegal}} \citep{Girardi2005},
and from the distribution of the SDSS spectroscopically confirmed stars detected also
in 2MASS \citep{Skrutskie2006}. Our colour cuts are a simplified version of those 
adopted by \citet{Richards2002} on the SDSS photometric database. We also check the 
consistency of the sources not classified as stars with the reference colours of 
Elliptical, Sc galaxies and QSOs obtained from the templates of \citet{Mannucci2001} 
and of \citet{Francis1991}. In summary in the $15\arcmin\times15\arcmin$ region explored 
we detect 2048 sources, 807 of which are classified as star on the basis of their colours.
This yields a number density of $\sim3.5$ stars per arcmin$^2$ that roughly correspond
to the prediction of the \texttt{Trilegal} software in this region of the sky
($\sim3.1$\,arcmin$^{-2}$).
The average surface density of extragalactic sources is $3.7\pm0.2$\,arcmin$^{-2}$, that is
consistent (within the uncertainties) with the $4.0\pm0.2$\,arcmin$^{-2}$ observed in the
The Great Observatories Origins Deep Survey \citep[GOODS,][]{Giavalisco2004} 
once we consider our \mbox{z--band} sensitivity limits.

\section[]{Discussion}\label{sec:3}

In this section we discuss the possible scenarios for this peculiar 
system.

\subsection{Gravitational lens}

\begin{figure}
\centering
\includegraphics[width=1.00\columnwidth]{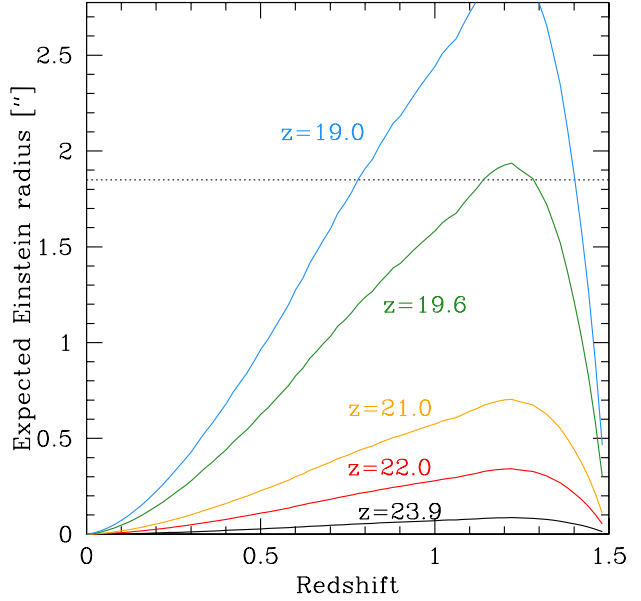}
\caption{Expected Einstein ring radius for a lensing galaxy with
various (observed) z--band magnitudes, as a function of redshift. 
The horizontal line shows the half--separation between \qqqb\ and~C.
The elliptical galaxy template by \citet{Mannucci2001} is used
in order to compute $k$-- and filter--corrections. We assume that
the potential well of the lensing galaxy is well described by
a SIS model with $\sigma=\sigma_*$, where $\sigma_*$ is the 
stellar velocity dispersion as set by the Faber-Jackson relation
(see text for details).
In the lensing scenario, a galaxy of z$\approx$19.6 would be 
required in order to justify the observed separation between 
\qqqb\ and~C. Thanks to the depth of our observations 
(z$_{\rm lim}\approx23.9$), we can rule out the 
strong--lensing scenario for these two sources.}\label{fig:3}
\end{figure}

To test the lens hypothesis for \qqq\ we consider two different
configurations: in case (i) \qqqb\ and \qqqc\ are images of the 
same, gravitationally lensed QSO while \qqqa\ is a different
QSO at similar redshift. In case (ii), also \qqqa\ is an image of 
the same QSO. We note that small differences are reported 
in the colours and in the spectra of the three QSOs (e.g., the \mgii\ 
lines peak at slightly different redshifts). This already disfavours 
the lensing scenario.
However, due to the limited signal--to--noise ratio of the available spectra,
we will focus our analysis mostly on the images.

In case (i) we model the potential well of the lensing object with 
a singular isothermal sphere (SIS). Multiple images of similar brightness
are obtained only along the Einstein ring, the radius of which is 
parametrised as $\theta_{\rm E}$ \citep[e.g.,][]{Narayan1990,Chieregato2007}:
\begin{equation}
\theta_{\rm E} = 4\pi\left(\frac{\sigma}{c}\right)^2 \frac{{\rm D}_{\rm LS}}{{\rm D}_{\rm S}}
\end{equation}
Here $\sigma$ is the velocity dispersion of the SIS, $c$ 
is the speed of light, ${\rm D}_{\rm LS}$ and ${\rm D}_{\rm S}$ are the 
angular diameter distances between the source and the lens, and between the 
source and the observer, respectively.
The minimum value of $\theta_{\rm E}$ allowed in this system corresponds
to half of the separation in the sky between \qqqb\ and C, i.e., $1\farcs85$.
This implies that, for a given redshift of the lens, we can infer a lower
limit for $\sigma$. If we adopt the working assumption that $\sigma=\sigma_*$
(the velocity dispersion of stars in the galaxy), we can use the 
Faber--Jackson relation \citep{Faber1976} to convert the lower limit on 
$\sigma$ into a luminosity limit for the lens galaxy. We use the updated
version of the Faber--Jackson relation reported by \citet{Nigoche2010},
\begin{equation} 
\log{\frac{\sigma_*}{\rm km\,s^{-1}}} = (-1.208\pm0.205) - (0.157\pm0.009) \, {\rm M}_{\rm r}
\end{equation}
where ${\rm M}_{\rm r}$ is the rest-frame r-band absolute magnitude. We convert
the observed z--band into ${\rm M}_{\rm r}$ using the filter and $k$--corrections
computed basing on the elliptical galaxy template by \citet{Mannucci2001}. 
Figure \ref{fig:3} shows that a galaxy brighter than z=19.6 is necessary to explain 
the observed geometry of the system in terms of strong gravitational lensing. 
The presence of such a galaxy is ruled out by our Omega2000 images.

Case (ii) is more extreme since wide separation lens systems are rare. In 
their search for lensed QSOs  with angular separation less than 
$\ad\lsim20\arcsec$, \citet{Inada2008} find in the SDSS only 3 out of 22 
systems with $\ad>3\arcsec$. \citet{Hennawi2007} 
estimated that in the SDSS QSO sample there should be only a few systems ($\lsim4$) 
multiply lensed by galaxy clusters with separations $>20\arcsec$. Moreover, triple imaged 
QSOs are not commonly observed \citep[e.g.,][]{Lawrence1984}.  

\begin{figure}
\centering
\includegraphics[width=1.00\columnwidth]{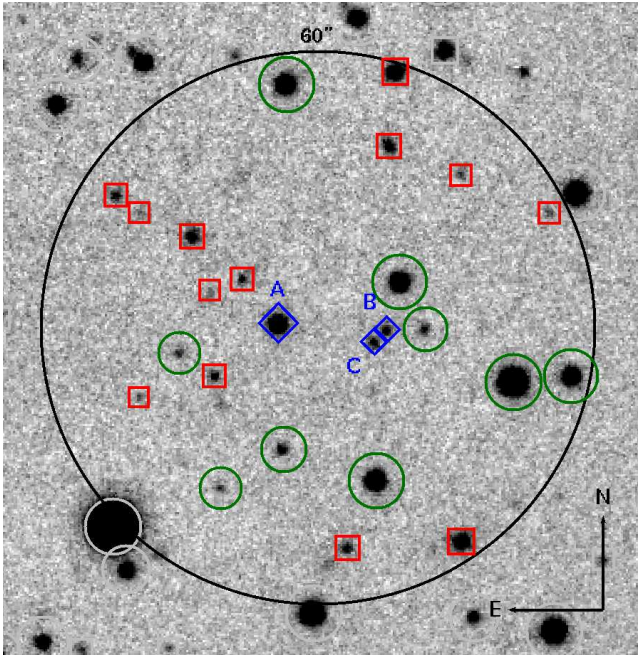}
\caption{Image of the $60\arcsec$ around the triplet \qqq\ as imaged in z--band at
the Calar Alto Observatory. Blue diamonds are the three QSOs. The 13 sources identified
as galaxies are shown as red squares, while the 9 stars as green circles. The surface
densities of Galactic and extragalactic sources in this area is consistent with
the detections in the rest of the image. 
}\label{fig:4}
\end{figure}

The three largest separation QSO lenses known so far, SDSS~J1029+2623 
\citep[$\Delta\theta=22\farcs5$, $\z=2.197$,][]{Inada2006},
SDSS~J2222+2745 \citep[$\Delta\theta=15\farcs1$, $\z=2.82$,][]{Dahle2012},
and SDSS~J1004+4112 \citep[$\Delta\theta=14\farcs6$, $\z=1.734$,][]{Inada2003}, 
have separation similar to the ones reported here. In all these cases
massive galaxy clusters act as lens \citep{Oguri2004, Inada2006, Dahle2012}. 
As a zeroth order test, we estimate the velocity dispersion required to 
explain the separation of the images assuming again a SIS profile for the
lens. As in case (i) we consider a limit for the Einstein radius 
$\theta_{\rm E}\approx10\arcsec$ that corresponds to $\sigma\approx600$\,\kms\
at $\z\approx0$ and $\sigma\approx1200$\,\kms\ at $\z\approx1$.
In our image the presence of such clusters should be apparent, but in fact
the surface density of galaxies within $60\arcsec$ from \qqq\
(in z--band $4.1\pm1.1$\,arcmin$^{-2}$) is consistent with those detected in
the whole frame ($3.7\pm0.2$\,arcmin$^{-2}$, see \S\ref{sec:2.2} and
Figure~\ref{fig:4}). 
For instance, we consider the case that in our field was present
XMMU~J100750.5+125818: a galaxy cluster located at redshift 
$\z\approx1.08$ with a line--of--sight velocity dispersion of 
$\sim600$\,\kms\ \citep[][]{Schwope2010}. If the cluster was centred
on \qqq\, in the first $60\arcsec$ there should be at least 10 galaxies 
in excess to the background. This is in contrast to what observed in 
the z--band where we detect only 12 galaxies instead of 22.
We note that this estimate is conservative since we considered
as reference only the (few) spectroscopically confirmed cluster members of
XMMU~J100750.5+125818, while the cluster should be more populated.

From the analysis of photometric data, we thus conclude that \qqq\ is not 
a case of gravitation lens.

\subsection{Chance superposition}

\qqq\ is the second triplet of QSOs at similar redshift known, after
QQQ~J1432-0106 \citep{Djorgovski2007}. {In Section~\ref{sec:2} we have 
already shown that these are extremely rare objects.} Here we estimate 
the expected number of QSO triplets in our sample that are originated 
by chance superposition of otherwise (physically) disconnected QSOs.
We start from the samples of spectroscopically confirmed QSOs 
by \citet{Schneider2010} and the sample of photometrically--selected
QSO candidates by \citet{Richards2009} (both based on SDSS data)
and we rely upon a procedure similar to the redshift permutation
method \citep[e.g.,][]{Osmer1981, Zhdanov2001}.
We assume that the QSOs have the same position in the sky as in the 
original samples but a random redshift is assigned through a Monte Carlo 
simulation that takes into account the redshift distribution of our 
complete sample ($\sim940000$ objects). We find that $\sim 0.05$ QSO 
triplets with $\dv<2000$\,\kms\ and projected separations 
$<200$\,kpc are expected. This result is an upper limit for the number 
of chance triple QSOs since in our computation most of the correlation
was destroyed, but we kept the angular distribution of the QSOs as in 
the original sample.  

\subsection{A common physical structure}\label{sec:3.3}

\begin{figure}
\centering
\includegraphics[width=1.00\columnwidth]{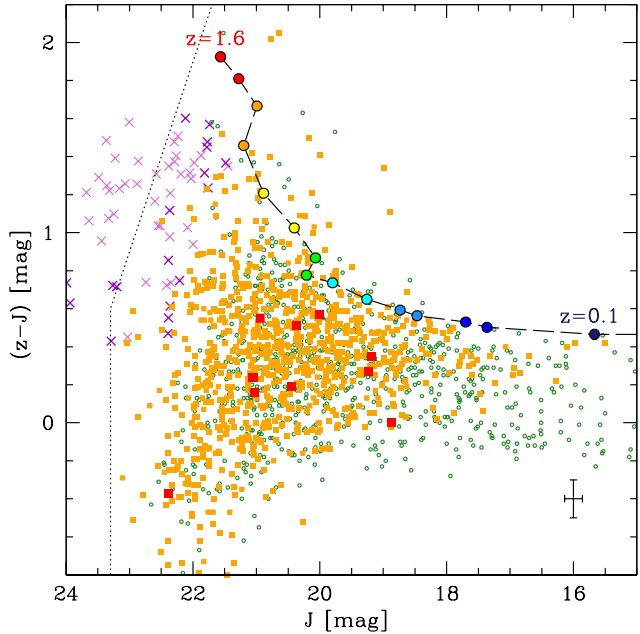}
\caption{Observed colour--magnitude diagram for the objects detected by 
\texttt{SExtractor} in the field of \qqq (green circles). Sources selected
as galaxies (see \S\ref{sec:2.2}) are shown in orange, and those that
lie within $60\arcsec$ from the triplet are highlighted with red squares.
The vertical dotted line marks the 5\,$\sigma$ detection limit
(note that \texttt{SExtractor} is on average more conservative
than the statistical limit magnitude). Typical error bars are shown
as a cross in the bottom--right of the figure. The dashed line
shows the location of an ${\rm M}^\star(\z)$ galaxy at various redshifts,
as derived from \citet{Ilbert2005}, and assuming the elliptical
galaxy template by \citet{Mannucci2001} for $k$--corrections.
For reference, we show with light violet crosses the location of the 
galaxies of the galaxy cluster XMMXCS~J2215.9-1738 \citep[$\z=1.46$, 
velocity dispersion $\sigma_v=580$\,\kms,][]{Stanford2006, Hilton2007, 
Hilton2009}. Dark violet crosses highlight the spectroscopic confirmed 
associations.
We do not find any evidence of an overdensity of galaxies with
colours consistent with a red sequence at $z=1.51$, thus suggesting 
that the environment of \qqq\ is not a rich cluster. 
}\label{fig:5}
\end{figure}

\begin{figure}
\centering
\includegraphics[width=1.00\columnwidth]{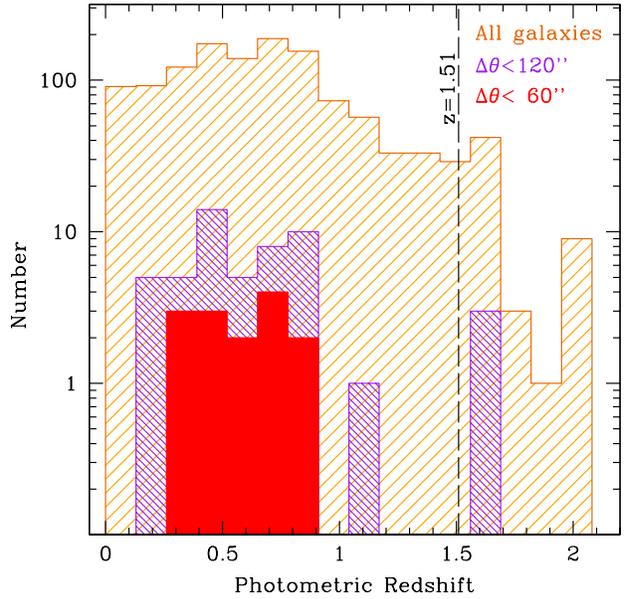}
\caption{
Distribution of the photometric redshifts estimated with \texttt{HyperZ} for 
the galaxies selected in the field of \qqq\ (orange histogram, see 
\S\ref{sec:3.3}). 
Sources that reside within $120\arcsec$ and $60\arcsec$ from the 
triplet are shown in purple and red, respectively. The vertical dashed
line marks the quasar triplet redshift. Typical errors associated to the 
photometric redshift determination are $\sim0.2$. None of the galaxies close 
to \qqq\ has a SED consistent with that of a galaxy at $\z\sim1.5$, confirming 
that there is no strong galaxy overdensity around the quasar 
triplet.}\label{fig:6}
\end{figure}

In the hypothesis that the system is virialised, the velocity differences 
among \qqqa, \qqqb\ and \qqqc\ (estimated from the broad \mgii\ emission 
line redshifts) imply a dynamical mass of $\sim10^{13}\msun$,
i.e. the system would lie inside a massive structure. In order to test
this scenario, we search for galaxies at redshift $\sim1.5$ in the
proximity of the system.
In Figure~\ref{fig:5} we show the observed colour--magnitude
diagram of the sources in our field. We compare the photometry of 
these sources with the apparent magnitude and colours of a 
${\rm M}^\star(\z)$ galaxy, where ${\rm M}^\star(\z)$ is
the characteristic luminosity (in absolute magnitudes) of galaxies
at a given redshift $\z$ in the rest--frame B--band, as computed by
\citet{Ilbert2005}. For filter and $k$--correction, we refer to the
elliptical galaxy template by \citet{Mannucci2001}. 
We do not find evidence of a strong overdensity of sources with colours 
consistent with those of a red sequence at $\z\approx1.5$. This suggests 
that \qqq\ is not located in a rich galaxy cluster or that the red 
sequence is not yet formed. We also estimate the photometric
redshifts of the sources in the field of \qqq\ with 
\texttt{HyperZ}\footnote{\texttt{http://webast.ast.obs-mip.fr/hyperz/}} 
\citep{Bolzonella2000} relying upon different galaxy templates (i.e.,
starburst, elliptical, lenticular, different type of spirals, and irregular
galaxies) and assuming the extinction law by \citet[][]{Calzetti2000}. 
It is worth noting that this 
choice of dust extinction model has only marginal effects on our results. 
This analysis confirms that none of the objects detected within $60\arcsec$
(that corresponds to $\sim500$\,kpc at the triplet redshift) and 
$120\arcsec$ ($\sim1$\,Mpc) have colours consistent with a galaxy at 
$\z\sim1.5$ (see Figure~\ref{fig:6}).
We thus propose that the physical structure, where the three QSOs reside, 
may be caught in the act of formation \citep[similarly to what reported in 
e.g.:][]{Decarli2009, Decarli2010a, Farina2011}. 

\section[]{Conclusions}\label{sec:4}

We report the discovery of a triplet of QSOs with angular separations
between $3\farcs7$ and $23\farcs5$ at a similar redshift of $\z\approx1.5$. 
The three sources have relative velocities within $ \sim 1000$\,\kms\ as 
derived from our measurements of the \mgii\ broad emission line and of 
the surrounding \feii\ multiplets in the observed optical spectra. 
From the analysis of u, g, r, i, z, J, and H broad band images we conclude 
that this system consist of three distinct sources rather than being a 
multiply imaged lensed QSO. This would be the second triplet of QSOs 
discovered so far after QQQ~J1432-0106 \citep{Djorgovski2007}.
We estimate that the probability that these systems are due to chance 
superposition are negligible. No clear galaxy overdensity is reported at 
photometric redshift consistent with the one of the triplet. 
We therefore propose that these systems are part of an ongoing common 
physical structure formation. 

The projected separation of \qqqa\ from the other two members of the system 
is greater than the typical distances between interacting systems 
\citep[$\sim50$\,kpc; e.g.,][]{Hennawi2006, Foreman2009}. This implies that 
the observed nuclear activity of \qqqa\ was probably not triggered during an 
interaction involving the three systems at the same time. On the other hand, 
due to the smaller separation observed between \qqqb\ and C, the triggering 
of nuclear activity by galactic interactions may be a viable scenario for 
these QSOs.
More conclusive constraints on the dynamics of this peculiar system and
on the properties of its environment will be obtained from future
higher signal--to--noise ratio spectroscopic data in the NIR band and deeper 
photometric images. 

\section*{Acknowledgements}

We thank the Calar Alto staff for the generous allocation of DDT time
and the prompt execution of our programme.
We acknowledge A.~Treves for helpful suggestions and comments on the
manuscript. We would also like to acknowledge the support and the 
advertising given by M.~Colpi, G.~Gavazzi, and F.~Haardt.
For this work EPF was supported by Societ{\`a} Carlo Gavazzi S.p.A. and by 
Thales Alenia Space Italia S.p.A. 
RD acknowledges funding from Germany's national research centre for 
aeronautics and space (DLR, project FKZ 50 OR 1104).
Support for MF was provided by NASA through Hubble Fellowship grant 
HF-51305.01-A awarded by the Space Telescope Science Institute, which is 
operated by the Association of Universities for Research in Astronomy, Inc., 
for NASA, under contract NAS 5-26555.
For this work we use: 
(i) spectra from ESO/NTT Telescope in La Silla;
(ii) imaging from CAHA/3.5m Telescope; and
(iii) data from the Sloan Digital Sky Survey. 
Funding for SDSS-III has been provided by the Alfred P. Sloan Foundation, the Participating Institutions, 
the National Science Foundation, and the U.S. Department of Energy Office of Science. The SDSS-III web 
site is {\texttt http://www.sdss3.org/}.
SDSS-III is managed by the Astrophysical Research Consortium for the Participating Institutions of 
the SDSS-III Collaboration including the University of Arizona, the Brazilian Participation Group, 
Brookhaven National Laboratory, University of Cambridge, University of Florida, the French Participation 
Group, the German Participation Group, the Instituto de Astrofisica de Canarias, the Michigan State/Notre 
Dame/JINA Participation Group, Johns Hopkins University, Lawrence Berkeley National Laboratory, Max Planck 
Institute for Astrophysics, New Mexico State University, New York University, Ohio State University, 
Pennsylvania State University, University of Portsmouth, Princeton University, the Spanish Participation 
Group, University of Tokyo, University of Utah, Vanderbilt University, University of Virginia, University of 
Washington, and Yale University.

\label{lastpage}

\end{document}